# Topological defect states in elastic phononic plates

Baizhan Xia[1*], Liang Tong[1], Jie Zhang[1], Shengjie Zheng[1], Xianfeng Man[2]

1 State Key Laboratory of Advanced Design and Manufacturing for Vehicle Body, Hunan University, Changsha, Hunan, People's Republic of China, 410082

2 College of Mechanical and Electrical Engineering, Changsha University, Changsha, Hunan, People's Republic of China, 410022

Topological defects (including disclinations and dislocations) which commonly exist in various materials have shown an amazing ability to produce excellent mechanical and physical properties of matters. In this paper, disclinations and dislocations are firstly introduced into the valley-polarized elastic phononic plate. Deformation of the lattice yields the interface expressing as the topologically protected wave guiding, due to the valley-polarized phase transition of phononic crystals (PnCs) across the interface. Then, disclinations are introduced into the Wannier-type elastic phononic plate. The deformation of the lattice yielded by disclinations produces a pentagonal core with the local five-fold symmetry. The topological bound states are well localized around the boundaries of the pentagonal cores with and without the hollow regions. The topological interface state and the topological bound state immunize against the finite sizes and the moderate disturbances of plates, essentially differing from the trivial defect states. The discovery of topological defect states unveils a new horizon in topological mechanics and physics, and it provides a novel platform to implement large-scale elastic devices with robust topological waveguides and resonators.

---

[*] xiabz2013@hnu.edu.cn

# 1 Introduction

Phononic crystals (PnCs) artificially designed in periodic elastic lattices have exhibited amazing properties which are usually impossible to be observed in natural materials, such as the negative refraction(Dong, Zhao et al. 2017, Tallarico, Movchan et al. 2017, Sridhar, Liu et al. 2018), the non-reciprocal propagation(Nassar, Xu et al. 2017, Zhao, Zhou et al. 2020), the wave trapping(Morvaridi, Carta et al. 2018), and the active cloaking(Darabi, Zareei et al. 2018, Meirbekova and Brun 2020, Zhang, Chen et al. 2020). Recently, inspired by topological insulators (TIs) discovered in condensed matters, topological states of elastic waves against defects and disorders have also been developed in PnCs(Chen, Nassar et al. 2018, Chen, Liu et al. 2019, Nanthakumar, Zhuang et al. 2019, Zhou, Wu et al. 2020). Topological PnCs can completely stop the wave propagation in their bulks, while robustly enable the wave propagation along well-defined interfaces or boundaries, providing excellent frameworks for the robust control of elastic waves, from wave guiding, beam splitting to frequency converting. Up to now, there are three main types of topological PnCs which are respectively developed by analogies of the quantum hall effect (QHE)(Khanikaev, Fleury et al. 2015, Nash, Kleckner et al. 2015, Wang, Lu et al. 2015, Yang, Gao et al. 2015, Souslov, van Zuiden et al. 2017, Mitchell, Nash et al. 2018), the quantum spin hall effect (QHE)(Mousavi, Khanikaev et al. 2015, He, Ni et al. 2016, Xia, Liu et al. 2017, Foehr, Bilal et al. 2018, Yu, He et al. 2018, Zheng, Theocharis et al. 2018, Wang, Bonello et al. 2019) and the quantum valley hall effect (QHE)(Lu, Qiu et al. 2017, Vila, Pal et al. 2017, Cha, Kim et al. 2018, Liu and Semperlotti 2018, Yan, Lu et al. 2018, Zhu, Liu et al. 2018, Liu and Semperlotti 2019, Ma, Sun et al. 2019, Wang, Bonello et al. 2019). A new class of symmetry-protected topological phases, characterized by the bulk polarization but not the quantized topological invariance, has led to the production of higher-order topological states. Higher-order topological PnCs do not exhibit gapless edge states, but instead topological corner states on the "boundaries of boundaries" (Serra-Garcia, Peri et al. 2018, Chen, Xu et al. 2019, Fan, Xia et al. 2019, Ni, Weiner et al. 2019, Xue, Yang et al. 2019, Zhang, Wang et al. 2019, Zhang, Long et al. 2019, Lin, Wang et al. 2020, Wakao, Yoshida et al. 2020, Zhang, Hu et al. 2020). Both topological edge states and topological corner states are tied to the nontrivial topology of the bulk band structure in the momentum space, so they are robust against the real-space disturbances of periodic lattices.

Topological defects, including disclinations and dislocations, are common crystalline defects in various materials (Kim, Zhao et al. 2009, Li, Cai et al. 2009) and are usually responsible for various mechanical and physical properties of matters (Lusk and Carr 2008, Wei, Wu et al. 2012, Wu and Wei 2013, Zhang, Li et al. 2014). In a hexagonal lattice, the disclination resulting from an extra (or missing) 60º crystalline wedge disrupts the orientational order and appears as a local five-fold (or seven-fold) symmetry with a topological charge of $-2\pi/6$ (or $+2\pi/6$)(Irvine, Vitelli et al. 2010). The dislocation which appears as the heptagon-pentagon dipole also disrupts the translational order. Topological defects in honeycomb lattices act on Dirac cone states with synthetic gauge fluxes, yielding topologically nontrivial states. Very recently, topologically-protected edge states and disclination states yielded by topological defects have been observed in photonic crystals consisting of macroscopic dielectric structures(Liu, Leung et al. 2020, Wang, Xue et al. 2020).

Topological defects provide a new platform for the realization of topological characteristics of classical wave systems beyond the general bulk-edge correspondence principle. However, up to now, topological features induced by topological defects have not been explored in elastic PnCs. In this paper, topological states yielded by topological defects will be systematically investigated in elastic phononic plates. Firstly, introducing topological defects into the valley-polarized elastic phononic plate, PnCs switch their valley-polarized phases across the interface yielded from the deformation of the lattice, leading to the valley-controlled wave propagation. Differing from the general topological PnCs, the topological phase transition in our elastic phononic plate is yielded from the deformation of the lattice, but not the elaborate design of two distinct lattices with opposite topological charges. Subsequently, joining disclinations into the Wannier-type elastic phononic plate, the deformation of the lattice produces a pentagonal core with the local five-fold symmetry. The topological bound states are well localized around the boundaries of the pentagonal core, being robust against the finite sizes and the moderate imperfects of the plate. Thus, topological defects open a new pathway toward the topological modulation of elastic waves.

## 2 Valley-polarized interface states induced by topological defects in elastic phononic plates

### *2.1 Theoretical analysis of topological defects for valley-polarized interface states*

The elastic phononic plate lattice is fabricated by cutting hexagonal blocks from an acrylic panel with the material properties of density $\rho=1190 kg/m^3$, Poisson's ratio $v=0.35$, and Young's modulus $E=3.2GPa$, shown in Fig. 1(a). The thickness of the acrylic panel is $d=1.98mm$. The width and length of the acrylic beam are $w=5.02mm$ and $L=15mm$, respectively. Two cylindrical nickel-plated neodymium magnets ($\rho=7400kg/m^3$, $v=0.28$, and $E=41GPa$), marked by light blue in Fig. 1a, are attached to the upper and lower sides of sublattices A and B, working as additional masses. The height and radio of the magnet are $h=2.0mm$ and $r=2.51mm$, respectively. Under a long wavelength limit, this lattice can be approximated as a thin plate whose out-of-plane mode is loosely coupled with in-plane modes. In this paper, the in-plane modes which go beyond the scope of this work are neglected.

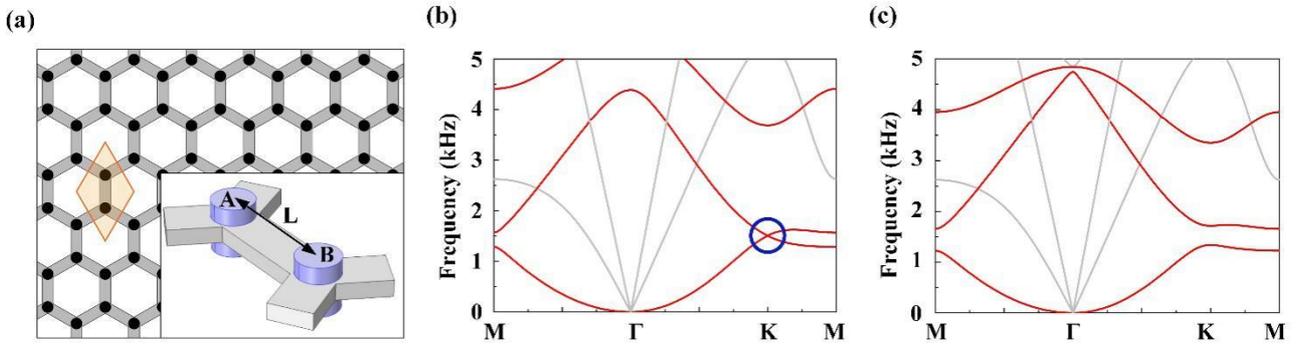

Fig. 1. (a). The elastic phononic plate with a unit cell marked by orange. (b) Band structure with a Dirac cone at $m_A=m_B$. (c) Band structure with a gap lift by the sublattice symmetry breaking ($m_A \neq m_B$).

The motion equation of elastic waves is a full vector wave equation, which can be expressed as

$$(\lambda + \mu)\nabla(\nabla \cdot \mathbf{u}) + \mu\nabla^2\mathbf{u} + \nabla\lambda(\nabla \cdot \mathbf{u}) + \nabla\mu \times (\nabla \times \mathbf{u}) + 2(\nabla\mu \cdot \nabla)\mathbf{u} = -\rho\omega^2\mathbf{u} \qquad (1)$$

in which $\mathbf{u}$, $\lambda$, $\mu$ and $\rho$ are functions of the spatial position $\mathbf{r}$. $\mathbf{u}(\mathbf{r})$ is the displacement vector field of elastic waves, $\lambda(\mathbf{r})$ and $\mu(\mathbf{r})$ are the independent elastic coefficients of elastic materials named as the Lamé constant, and $\rho(\mathbf{r})$ is the mass density of elastic materials, respectively.

In elastic materials with periodic structures, the propagation of elastic waves can be expressed as an intrinsic problem,

$$H(\mathbf{r})\mathbf{u_k}(\mathbf{r}) = \rho(\mathbf{r})\omega_\mathbf{k}^2\mathbf{u_k}(\mathbf{r}) \qquad (2)$$

in which $H$ is the Hamiltonian operator that can be expressed as,

$$H = -(\lambda + \mu)\nabla(\nabla) + \mu\nabla^2 + \nabla\lambda(\nabla \cdot) + \nabla\mu \times (\nabla \times) + 2(\nabla\mu \cdot \nabla) \qquad (3)$$

corresponding to the Bloch wave function $\mathbf{u_k}(\mathbf{r})$.

If the degeneracy at the point **k** is $d$, the corresponding eigenwave function is $\{\mathbf{u}_j(\mathbf{r})\}$. According to the $\mathbf{k} \cdot \mathbf{p}$ perturbation, the Bloch state $\mathbf{u}_{\mathbf{k}'}(\mathbf{r})$ of the point $\mathbf{k}'$ near the degenerate point **k** can be produced by the linear expansion of $\{\mathbf{u}_j(\mathbf{r})\}$, that is,

$$\mathbf{u}_{\mathbf{k}'}(\mathbf{r}) = e^{-i\Delta\mathbf{k}\cdot\mathbf{r}} \sum_{j=1}^{d} A_j \mathbf{u}_j(\mathbf{r}) \tag{4}$$

Substituting $\mathbf{u}_{\mathbf{k}'}(\mathbf{r})$ into the elastic wave equation and considering the orthogonality of the basis vector $\{\mathbf{u}_j(\mathbf{r})\}$, namely,

$$\int \mathbf{u}_i^*(\mathbf{r})\mathbf{u}_j(\mathbf{r})\rho(\mathbf{r})d\Omega = \delta_{ij} \tag{5}$$

we can obtain,

$$\sum_{j=1}^{d}\left[\omega_\mathbf{k}^2 \delta_{ij} + \Delta\mathbf{k}\cdot\mathbf{p}_{ij} + \Delta\mathbf{k}\Delta\mathbf{k}\!:\!q_{ij}\right] = \omega_{\mathbf{k}'}^2 A_i \tag{6}$$

$q_{ij}$ is a tensor, expressing as,

$$q_{ij} = \int\left[(\lambda+\mu)\mathbf{u}_i^*\mathbf{u}_j + \mu(\mathbf{u}_i^*\cdot\mathbf{u}_j)I\right]d\Omega \tag{7}$$

$\mathbf{p}_{ij}$ is a vector matrix, expressing as,

$$\mathbf{p}_{ij} = -i\int \Lambda_{ij}d\Omega \tag{8}$$

in which,

$$\Lambda_{ij} = (\lambda+\mu)\left(\mathbf{u}_i^*\nabla\cdot\mathbf{u}_j + \mathbf{u}_i^*\cdot\nabla\mathbf{u}_j\right) + 2\mu(\nabla\mathbf{u}_j)\cdot\mathbf{u}_i^* + \nabla\lambda\cdot\mathbf{u}_i^*\mathbf{u}_j + \mathbf{u}_i^*\mathbf{u}_j\cdot\nabla\mu + (\mathbf{u}_i^*\cdot\mathbf{u}_j)\nabla\mu \tag{9}$$

Without considering the higher-order terms and only taking the linear approximation of the perturbed Hamiltonian, Eq. (6) can be simplified as,

$$\sum_{j=1}^{d} H'_{ij} A_j = \omega_{\mathbf{k}'}^2 A_i \tag{10}$$

When the operation $R \in G_\mathbf{k}$ acts on the crystal, it means that the space coordinate **r** can be replaced by $R^{-1}\cdot\mathbf{r}$. For the eigenstate $\psi_i(\mathbf{r})$, $R$ is expressed as the function operator $O_R$, which changes the eigenstate to $O_R\psi_i(\mathbf{r}) = \psi_i(R^{-1}\cdot\mathbf{r})$. According to the group theory, eigenstate functions constitute a set of basic functions of $G_\mathbf{k}$ group. If the eigenstate is degenerated, we can obtain,

$$O_R\psi_i(\mathbf{r}) = \sum_{j=1}^{d} D_{R,ij}\psi_j(\mathbf{r}) \tag{11}$$

where $D_{R,ij}$ is a group representation of the operation $R$.

Under the operation $R$, the gradient operator $\nabla$ of the eigenstate is transformed to $R^{-1}\cdot\nabla$, while the material parameters of crystals remain unchanged due to the symmetry which is consistent with the crystal. In this case, $\mathbf{p}_{ij}$ can be rewritten as,

$$\mathbf{p}_{ij} = R^{-1}\sum_{m,n=1}^{d} D_{R,mi}^* D_{R,nj}^* \mathbf{p}_{mn} \tag{12}$$

Therefore, if the symmetries are the same, the same perturbed Hamiltonian can be obtained by group theory constraints. When the masses of A and B are same (namely $m_A=m_B$), a Dirac cone will degenerate at the high-symmetry point K of the Brillouin zone, shown in Fig. 1(b). When the sublattice symmetry is broken (namely $m_A \neq m_B$) by removing a magnet from the upper and lower sides of the sublattice B, the Dirac cone will be lift, shown in Fig. 1(c), yielding a valley-polarized state. By constructing an interface between two types of PnCs with the sublattice symmetry breaking (one is $m_A>m_B$ and the other one is $m_A<m_B$), the valley-polarized edge state will be produced(Lu, Qiu et al. 2017, Vila, Pal et al. 2017, Cha, Kim et al. 2018, Liu and Semperlotti 2018, Yan, Lu et al. 2018, Zhu, Liu et al. 2018, Liu and Semperlotti 2019, Ma, Sun et al. 2019, Wang, Bonello et al. 2019).

Deleting a 2π/6 sector and reattaching beams by joining sites of same sublattices, a pentagonal phononic plate with an interface yielded from the disclination of lattice is generated, as shown in Fig. 2(a). In the disclination of lattice, the lengths of beams are fixed. Fig. 2(a) shows that the sites of sublattices A and B switch each other across the interface. This is locally equivalent to having two decoupled valleys with opposite signs of on opposite sides. Taking a pair of sublattices A and B on the top of the interface, a pair of valleys with extreme frequency values are formed, as shown in Fig. 2(b). A pair of valleys are also formed at the bottom of the interface, as shown in Fig. 2(c). However, eigenstates of valleys at the bottom of the interface are reversed, when compared with those on the top of the interface. It indicates that the Dirac mass of each valley switches its sign across the interface.

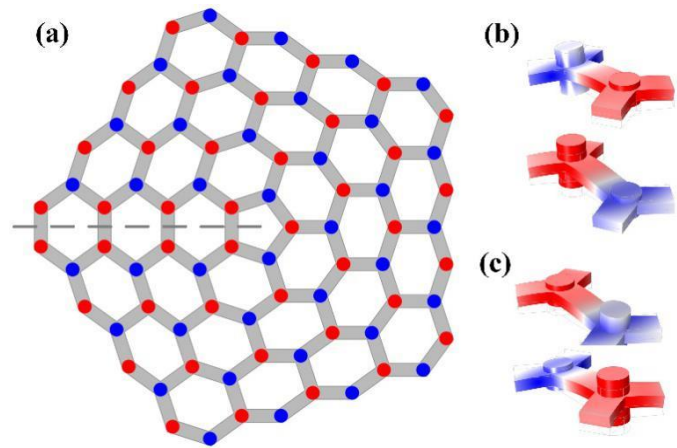

Fig 2. (a). Pentagonal elastic phononic plate. (b) Eigenstates of valleys on the top of the interface. (c) Eigenstates of valleys at the bottom of the interface.

The pentagonal elastic phononic plate can be designed by the following nearest-neighbor tight-binding Hamiltonian,

$$\mathcal{H} = -iv(\tau\sigma_1\partial'_1 + \sigma_2\partial'_2) - m\sigma_3 \tag{13}$$

Where $v$ denotes the Dirac velocity, $\tau$ is the valley Pauli matrix, $\sigma_i$ is the sublattice Pauli matrix and $m$ represents the sublattice detuning. $\partial'_1$ and $\partial'_2$ are distortions of sublattices which can be modeled by a local frame rotation,

$$\begin{pmatrix}\partial'_1\\\partial'_2\end{pmatrix} = \begin{pmatrix}cos\Theta(r) & -sin\Theta(r)\\sin\Theta(r) & cos\Theta(r)\end{pmatrix}\begin{pmatrix}\partial_1\\\partial_2\end{pmatrix} \tag{14}$$

where $\Theta(r)$ is a position-dependent frame rotation angle which is discontinuous across the disclination. The Hamiltonian equation with topological defects is similar to the general one without topological defects. In this case, the disclination yields a valley-polarized interface consisting of strings of nearest-neighbor sublattices with opposite valley-polarized states. The general valley-polarized interface must be a loop or throughout the whole plate, while the disclination can optionally terminate at any location of the bulk, providing an additional degree-of-freedom to design topologically protected interfaces.

## *2.2 Valley-polarized interface states induced by topological defects*

### *2.2.1 Eigenstates of valley-polarized elastic phononic plates with disclinations*

To verify eigenstates localized along the interface, the full-wave numerical simulation for the eigenvalue equation of the pentagonal elastic phononic plate is performed. Fixed boundary conditions are applied at the boundaries of the plate. As shown in Fig. 3(a), eigenstates which are localized to the interface are observed in the band gap. Two simulated displacement field profiles for eigenstates, one for the eigenstate of disclination and the other one for the eigenstate of bulk, are shown in Figs. 3(c) and 3(d). For the first one, the elastic wave energy focuses along the interface. For the second one, the elastic wave energy scatters into the bulk of the plate. For comparison, we also investigated a hexagonal elastic phononic plate without topological defects. Eigenfrequencies near the bandgap are considered, as shown in Fig. 3(b). In such plate, there are no interface states in the band gap.

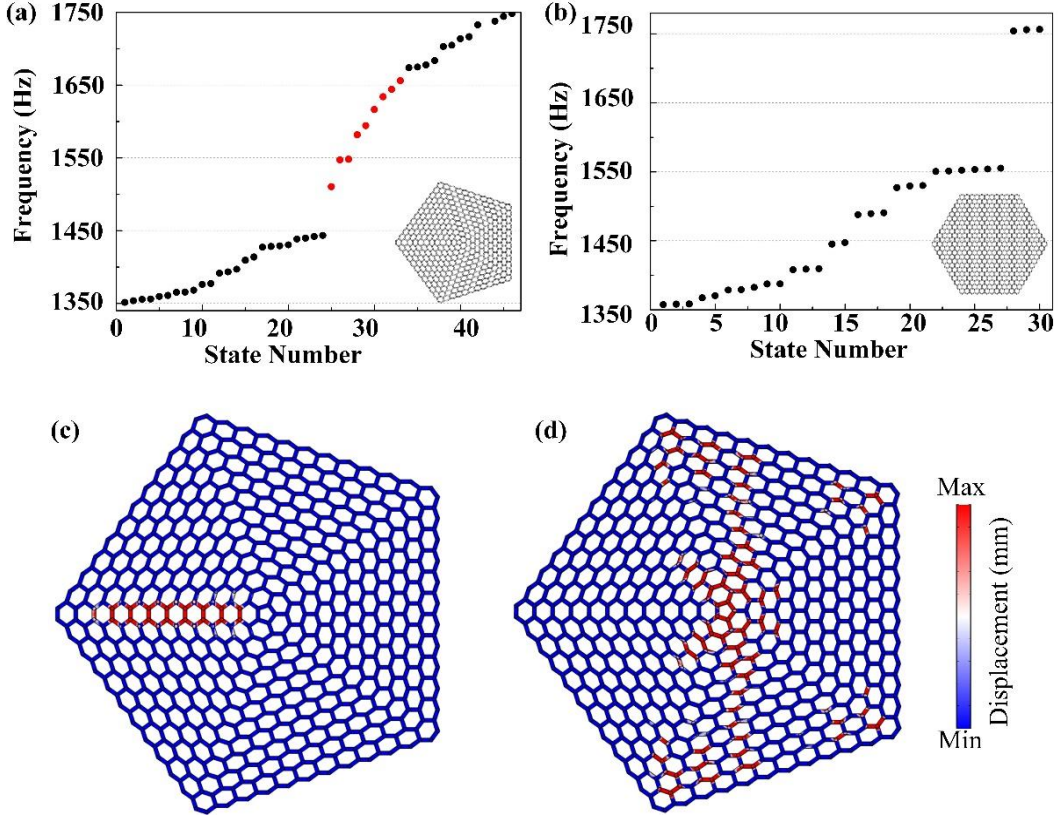

Fig. 3 (a)-(b) Eigenfrequency spectra of the pentagonal elastic phononic plate and the hexagonal elastic phononic plate. (c)-(d) The simulated displacement field profiles for eigenmodes localized on the disclination and the bulk of the pentagonal elastic phononic plate.

To investigate the topological defects acting on the elastic phononic plate without sublattice symmetry breaking, we implement a sample whose sublattices have equal masses, namely $m_A=m_B$, as shown in Fig. 4(a). The eigenfrequency spectrum of this sample are presented in Fig. 4(b). As expected, there is no indication of any gap near the Dirac frequency, predicted to be 1517Hz. This is consistent with the band structure of a unit cell, which possesses a Dirac cone but not a band gap. We examine all individual eigenstates from 1350Hz to 1750Hz and present an intensity distribution of eigenstate (near the Dirac cone) in Fig. 4(c). We found that the elastic wave energy is not localized at the interface yielded by topological defects. This is because the elastic phononic plate without sublattice symmetry breaking act on Dirac cone like a gauge field rather than a confining potential.

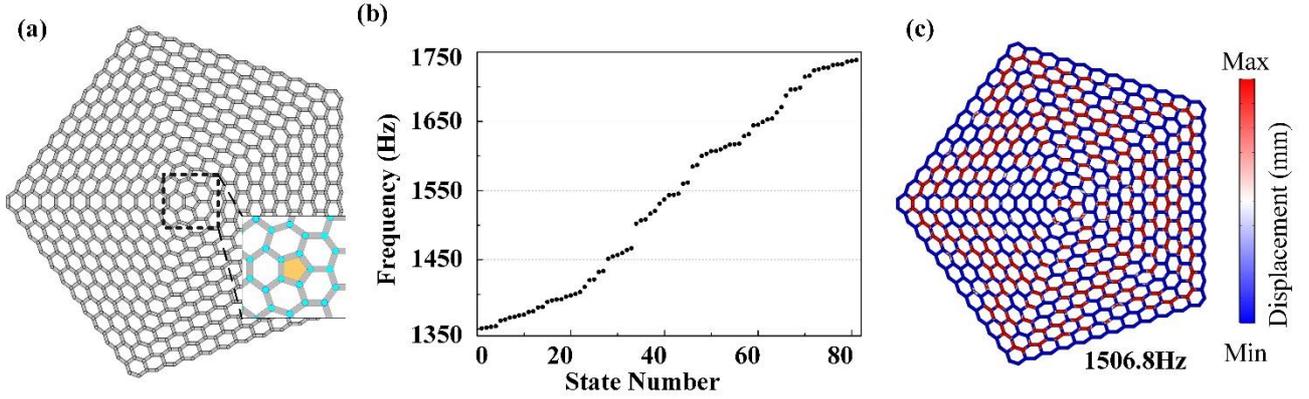

FIG. 4. (a) The pentagonal elastic phononic plate whose sublattice symmetry is not broken. Inset: schematic of the pentagonal defect at the center. (b) Eigenfrequency spectrum of the pentagonal elastic phononic plate. (c) The simulated displacement field profiles for the eigenstate near the Dirac cone.

## *2.2.2 Waveguiding induced by disclinations in valley-polarized elastic phononic plates*

The topological feature of valley-polarized interface states is the robustness of the wave propagation which immunes against disturbances, such as cavities and disorders. Cavities are produced by removing additional masses from several sublattices A and B in the bulk of the plate. Disorders are yielded by randomly shifting several sublattices in the bulk of the plate, yielding partially amorphous lattices. They are not the spin-mixing disturbances, thus the topological states along the interface will not be broken. The configurations of the pentagonal elastic phononic plates are presented in Fig. 5(a) for the one without disturbances, in Fig. 5(c) for the one with cavities and in Fig. 5(e) for the one with disorders. Interfaces are highlighted by orange domains. The cavities and disorders are marked by the red circles. Full-field numerical simulations are performed in the frequency domain under the time harmonic excitation at 1510Hz. Elastic waves are excited at topological defect cores, namely centers of pentagonal elastic phononic plates, marked by red stars. For the plate without disturbances, the elastic wave energy strongly concentrates along the interface and is well guided, with little leakage into the bulk, as shown in Figs. 5(b). Simulated displacement amplitude profiles in Figs. 5(d) and 5(f) further show that even though there are some cavities and disorders, the elastic waves can efficiently pass through interfaces without obviously attenuation.

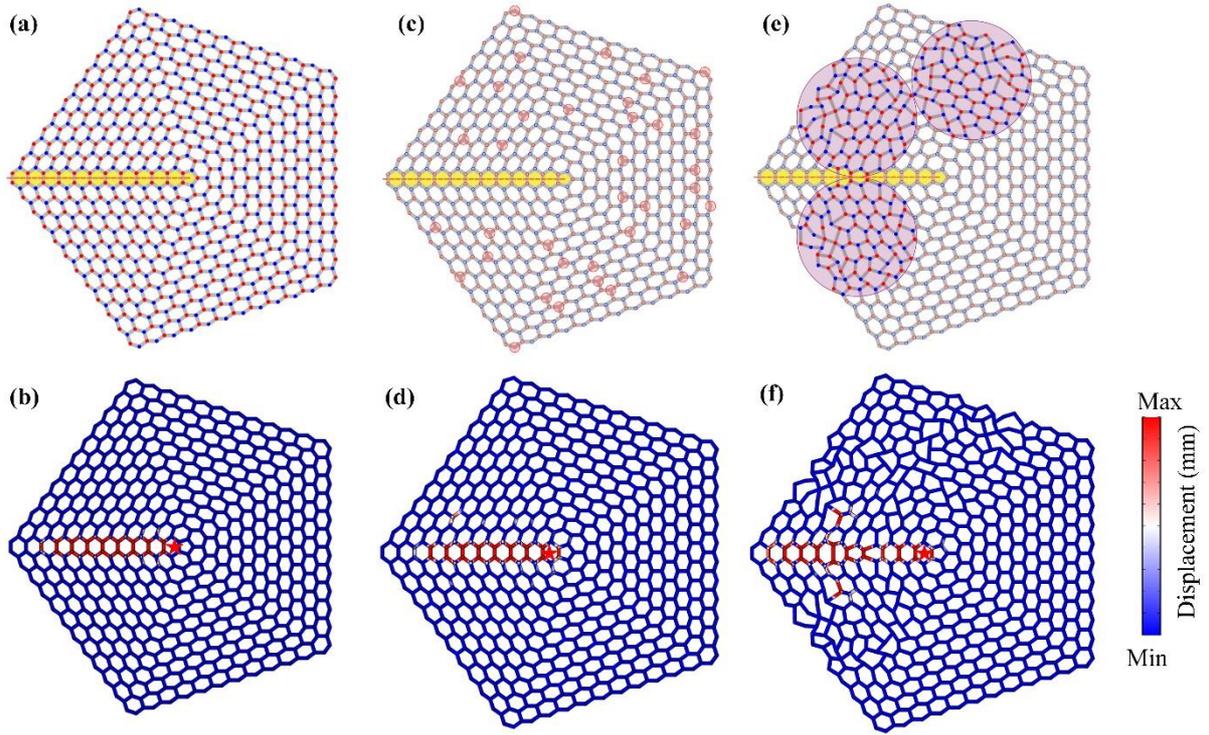

Fig. 5 The configurations of the pentagonal elastic phononic plates without disturbances (a), with cavities (c) and with disorders (e). The snapshots of displacement amplitudes of elastic waves for the pentagonal elastic phononic plates without disturbances (b), with cavities (d) and with disorders (f).

Following, we will systematically investigate the effects of structural disturbances (cavities and disorders) on the waveguiding of the plate. The structural disturbances originate from a sublattice and gradually seep into the surrounding areas. When the cavities are confined within the bulk of the plate away from the interface, as shown in Figs. 6(a)-6(c), the elastic waves can efficiently transport along the interface. As shown in Fig. 6 (d), when the cavities infiltrate into the interfaces, the elastic waves spread into the bulk of the plate, due to the serious breaking of the valley-polarized states at the bottom of the interface.

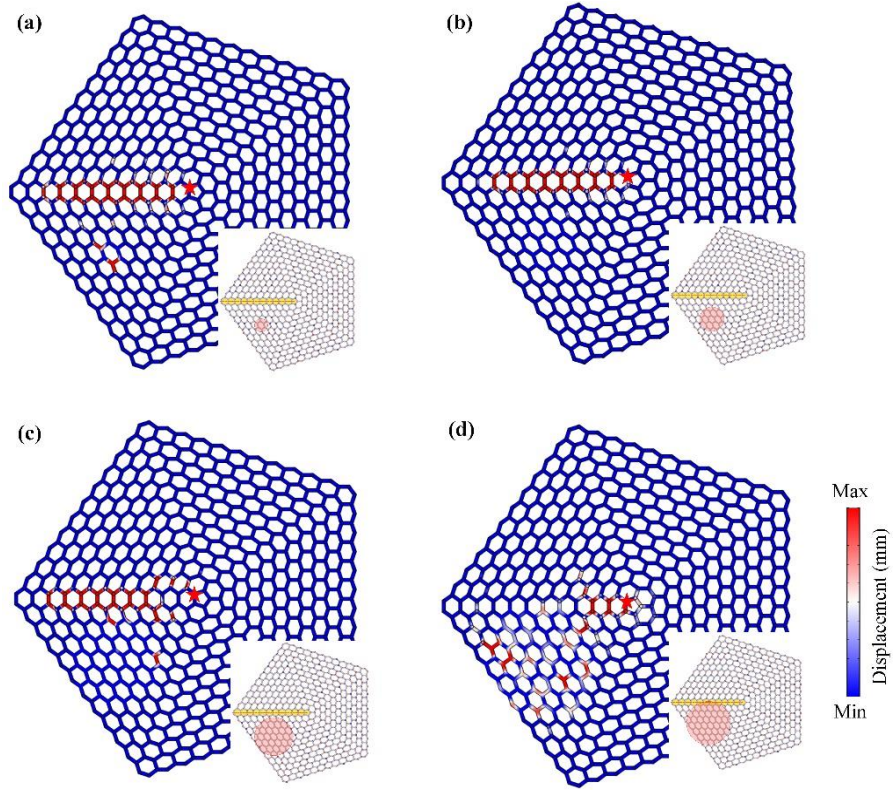

Fig. 6. The snapshots of displacement amplitudes of elastic waves for the pentagonal elastic phononic plates with cavities. (a)-(c) The cavities are confined within the bulk of the plate. (d) The cavities infiltrate into the interfaces.

When the disorders are confined within the bulk of the plate away from the interface, as shown in Figs. 7(a)-7(b), the elastic waves can efficiently transport along the interface. When the disorders infiltrate into the interfaces, the transmission efficiency of the elastic waves is still high along the waveguide, as shown in Fig. 7(c). Upon further increasing the radius of disorders, the interface states start to dissolve, resulting in that the elastic waves gradually spread into the bulk of the plate, as shown in Figs. 7(d). Therefore, the elastic phononic plate with topological defects is stable, at least when the interface of the plate is valley-polarized, despite the bulk being plagued by structural disturbances.

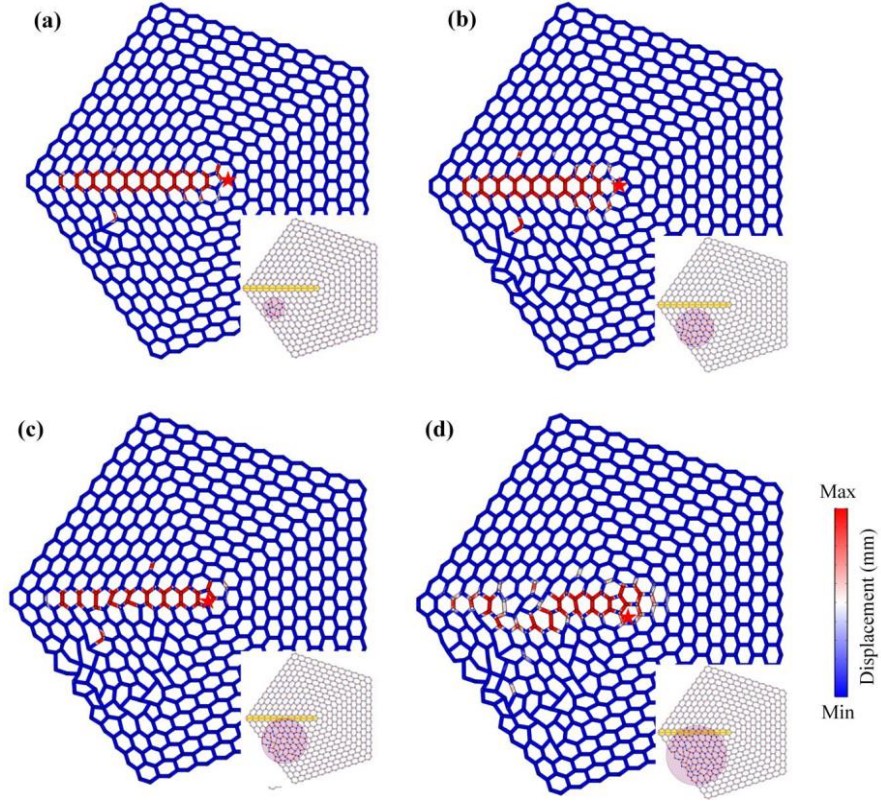

Fig. 7. The snapshots of displacement amplitudes of elastic waves for the pentagonal elastic phononic plates with disorders. (a)-(b) The disorders are confined within the bulk of the plate. (c) The disorders infiltrate into the interfaces. (d) The disorders pass through the interfaces.

## *2.2.3 Waveguiding induced by dislocations in valley-polarized elastic phononic plates*

Here, we design a topological interface induced by the dislocation, namely a sublattice moves from a hexagonal lattice to another one, as shown in Fig. 8(a). The hexagonal lattice with the missing sublattice works as the pentagonal topological defect, while the other one with the additional sublattice works as the heptagonal topological defects. It can be found from Fig. 8(a) that the sublattices A and B switch each other across the interface, leading to the transition of valley-polarized phases and guaranteeing the topologically protected interface propagation. For comparison, we design a conventional (nontopological) waveguide in the elastic phononic plate, by selectively removing additional masses along the desired route, shown in Fig. 8(b).

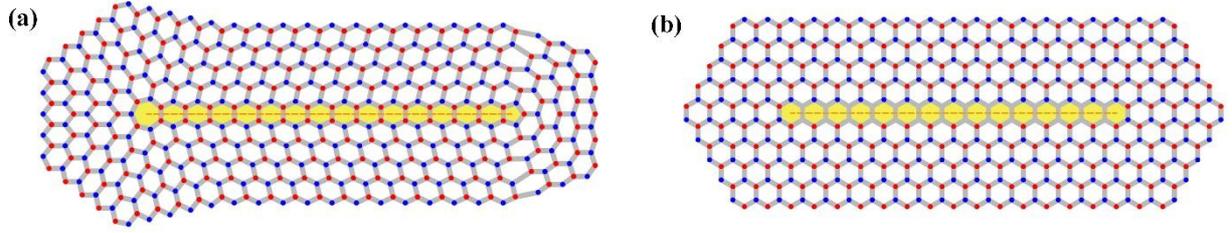

Fig. 8. (a) The elastic phononic plate with a topological interface. (b) The elastic phononic plate with a trivial waveguide.

In Fig. 9(a), we show full-wave simulations of the interface without disturbances and with cavities and disorders. The wave field distributions are excited by a 1550Hz source which is placed at the left end of the interface, marked by a red star. As shown in Fig. 9(a), the elastic wave energy can efficiently transport along the interface. When cavities and disorders are introduced in the plate, we found that the elastic wave energy can efficiently transport to the right end of the interface, exhibiting a good immunization against structural disturbances, as shown in Figs. 9(b) and 9(c). However, for the conventional waveguide, numerical simulations show that the elastic wave energy obviously scatters into the bulk of the plate, even though there are no disturbances, as shown in Fig. 9(d).

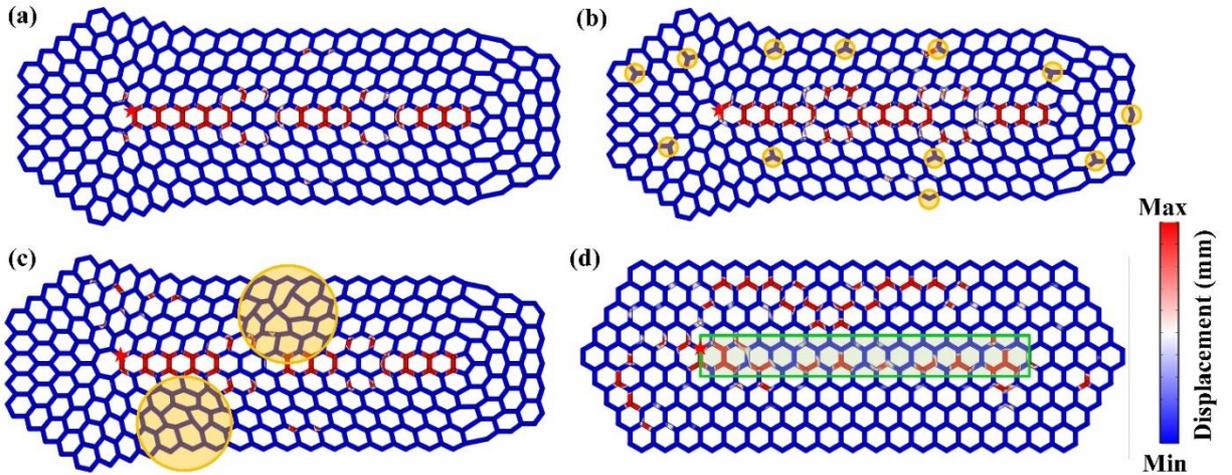

Fig. 9. Simulated field distributions for topological interfaces without disturbances(a), with cavities (b) and with disorders (c). (d) Simulated field distribution for the conventional waveguide without disturbances.

# 3 Topological bound states induced by disclinations in Wannier-type elastic phononic plates

### 3.1 Wannier centers of elastic phononic plates with pentagonal cores

We reconsider the elastic phononic plate, as shown in Fig. 10(a). Pseudospin-dependent edge states and higher-order corner states have been realized in this elastic phononic plate. Here, we will focus on the topological bound states induced by disclinations. The pentagonal elastic phononic plate can be constructed by removing a $2\pi/6$ sector and gluing the remaining nodes of the model. To be consistent with the definition of Wyckoff positions, we only remove an integral number of unit cells when constructing disclinations. The unit cell consisting of six nodes is shown in Fig. 10(b). The elastic beams linking the six nodes in a unit cell are the intra-cell beams whose coupling strengths are defined as the intra-cell coupling $1/l_{intra}$. The elastic beams connecting the nodes among the nearest-neighboring unit cells are the inter-cell beams whose coupling strengths are defined as the inter-cell coupling $1/l_{inter}$. Topological nontrivial ($l_{inter}>l_{intra}$, Fig. 11a) and trivial ($l_{inter}<l_{intra}$, Fig. 11b) elastic phononic plates have distinct responses to pentagonal cores, which can be defined by the fractional disclination charge extracted from the configuration of Wannier centers.

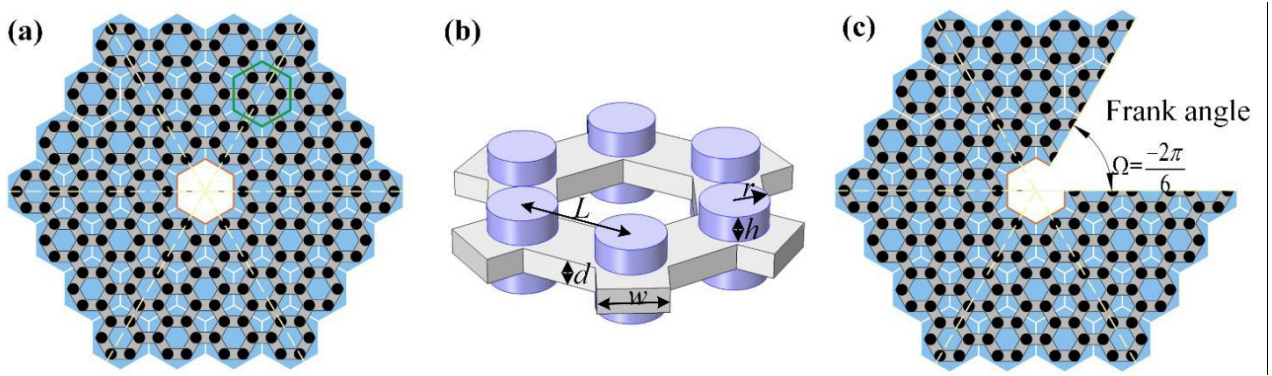

Fig 10. (a). The elastic phononic plate without topological defects. (b) The unit cell consisting of six nodes. Light blue cylinders are nickel-plated neodymium magnets working as additional masses. (c) The elastic phononic plate with a missing $2\pi/6$ crystalline wedge.

The topological nontrivial elastic phononic plate is a classification of $C_6$-symmetric TIs with $h_{3c}^{(6)}$, so its Wannier centers are located at the edges of unit cells (Benalcazar, Li et al. 2019, Li, Zhu et al. 2020), marked by blue solid circles in Figs. 11(c). Each Wannier center is shared by two neighboring unit cells and contributes $\frac{1}{2}$ charge to each one of them (Benalcazar, Li et al. 2019). There are five boundaries around the pentagonal core, giving rise to an overall charge of $\frac{5}{2}$. Thus, the pentagonal core yielded by disclinations must trap a charge of $\frac{1}{2}$, when compared with the unit cell with six boundaries whose charge is 3. In general, the disclination will locally break the rotation and translation symmetry of lattice, slightly shifting Wannier centers from Wyckoff positions. The deviation of Wannier centers changes the charge distribution, leading to a small perturbation ε of the fractional disclination charge at the pentagonal core. Generally, the perturbation ε of charge is much less than $\frac{1}{2}$, thus the disclination charge localized at the core is still a fraction closely to $\frac{1}{2}$. In the trivial elastic phononic plate, the Wannier centers are located at the center of the unit cell (Benalcazar, Li et al. 2019, Li, Zhu et al. 2020), marked by red solid circles in Figs. 11(d). In this case, all unit cells have integer charges. Thus, the trivial elastic phononic plates cannot produce fractional disclination charges.

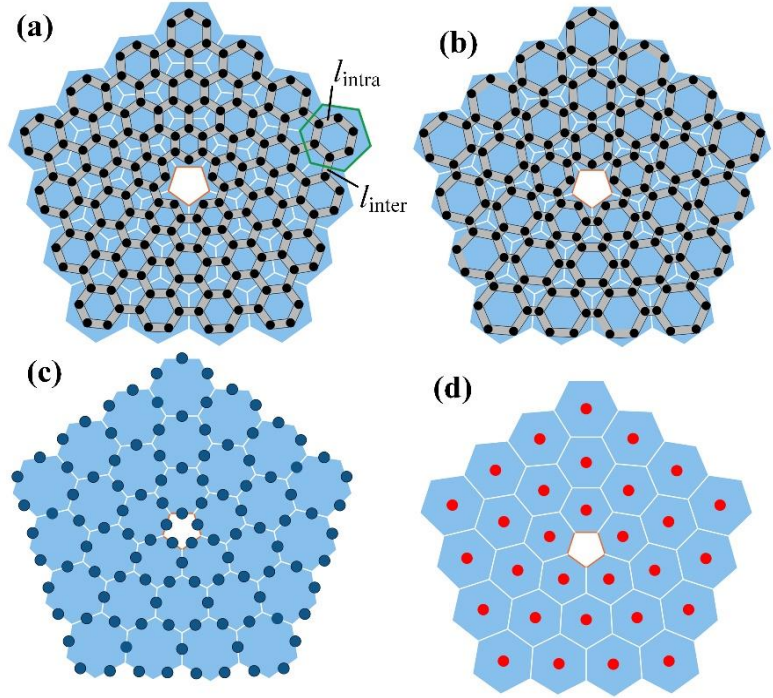

Fig. 11. Distributions of Wannier centers in elastic phononic plates with pentagonal cores. (a) and (b) Topological nontrivial and trivial elastic phononic plates with pentagonal cores. There are no beams in the hollow region of the plate. (c) For the topological nontrivial one, $l_{inter}$>$l_{intra}$, the Wannier centers marked by blue solid circles are located at the edges of unit cells. (d) For the trivial one, $l_{inter}$<$l_{intra}$, the Wannier centers marked by red solid circles are located at the centers of unit cells.

In order to calculate the quantized fractional disclination charge, an indice χ(6) for $C_6$-symmetry is given by (Benalcazar, Li et al. 2019)

$$\chi^{(6)} = \left[ M_1^{(2)}, K_1^{(3)} \right] \tag{15}$$

where $M_1^{(2)}$ and $K_1^{(3)}$ are the rotation invariances of the $C_6$-symmetry insulator. $M_1^{(2)} = 2$ and $K_1^{(3)} = 0$ for the nontrivial $C_6$-symmetric lattice with $h_{3c}^{(6)}$ (Benalcazar, Li et al. 2019).

The fractional charge at the pentagonal core yielded by disclinations in a $C_6$-symmetric insulator can be defined as (Benalcazar, Li et al. 2019, Li, Zhu et al. 2020):

$$Q_{dis} = \frac{\Omega}{2\pi}\left(\frac{3}{2}M_1^{(2)} + 2K_1^{(3)}\right) \bmod 1 \tag{16}$$

where $\Omega = \frac{2\pi}{6}$ is the Frank angle, the angle of removing sector. As $M_1^{(2)} = 2$ and $K_1^{(3)} = 0$, we can obtain $Q_{dis} = \frac{1}{2}$. The nonzero disclination index in the elastic phononic plate will indicate the generation of the topological bound states around the pentagonal cores.

*2.2 Bound states of elastic phononic plates with hollow pentagonal cores*

A topological nontrivial elastic phononic plate consisting of thirty-unit cells is considered. In full-wave numerical simulations of eigenstates, the free boundary conditions are applied at the boundaries of the plate. Eigenfrequencies near the bandgap are investigated. Fig. 12(a) shows that the bulk eigenmodes are truncated by a band gap. There are five topological bound states in the band gap, which are perfectly isolated from bulk states. Intensity distributions of topological bound states depicted in Figs. 12(b)-12(f) show that the elastic waves are strongly concentrated around the boundaries of the pentagonal core. These topological bound states are insensitive to the finite size of the plate. When the number of unit cells is decreased from 30 to 15 by removing the outermost unit cells, there are also five topological bound states in the band gap, as shown in Fig. 13. When the number of unit cell is increased from 30 to 50 by adding a row of unit cells on the periphery of the outermost unit cells, there are six bound states in the band gap, as shown in Fig. 14. Five of six bound states are topological ones, while the new one is a trivial bound state which has been found in the PnC with a hollow region. The regular hexagonal unit cell without deformation cannot spread out a fan-shaped area with an angle of $2\pi/5$. With the increase of the finite size, the outmost unit cell will become much larger than the innermost one. In this case, the trivial bulk states produced by the serious deformation of the plate will gradually infiltrate into the bulk band gap of the plate and the bulk band gap will be greatly shrunk, as shown from Fig. 13 to Fig. 12, and to Fig. 14.

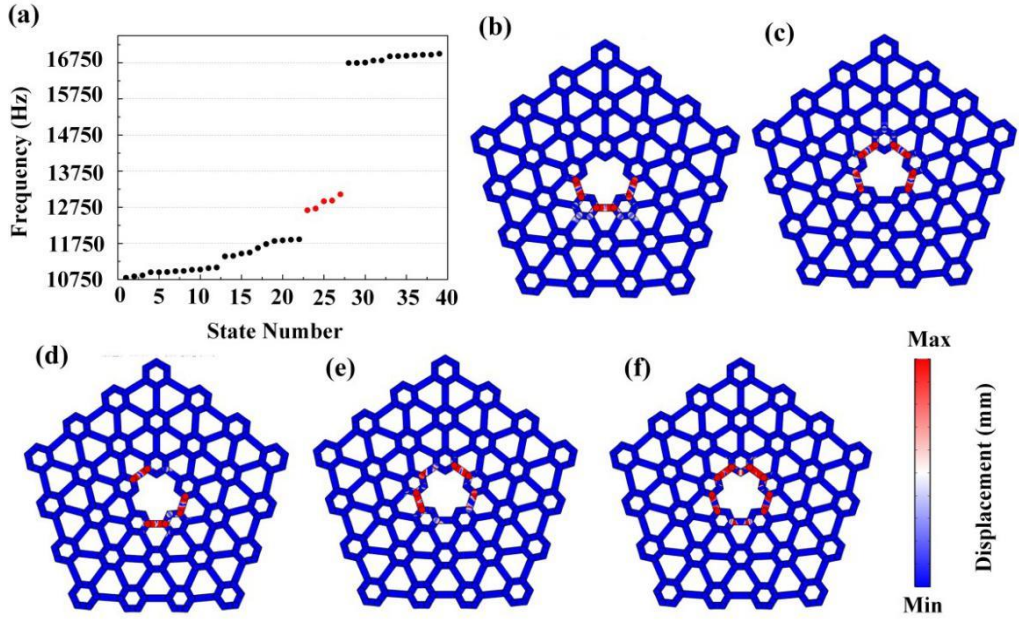

Fig. 12 (a) Eigenfrequency spectrum of the topologically nontrivial elastic phononic plate with a pentagonal core yielded by disclinations. The number of unit cells is 30. The red solid circles are the topological bound states. The black solid circles are the bulk states. (b)-(f) The simulated displacement field profiles of five topological bound states localized around the pentagonal core.

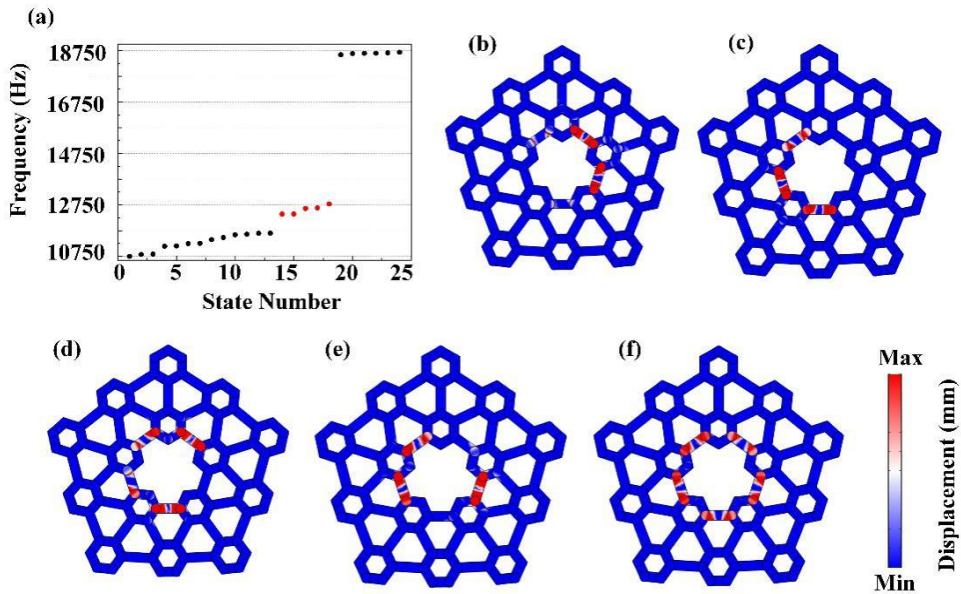

Fig. 13 (a) Eigenfrequency spectrum of the topologically nontrivial elastic phononic plate with the pentagonal core yielded by disclinations. The number of unit cells is 15. The red solid circles are the topological bound states. The black solid circles are the bulk states. (b)-(f) The simulated displacement field profiles of five topological bound states localized around the pentagonal core.

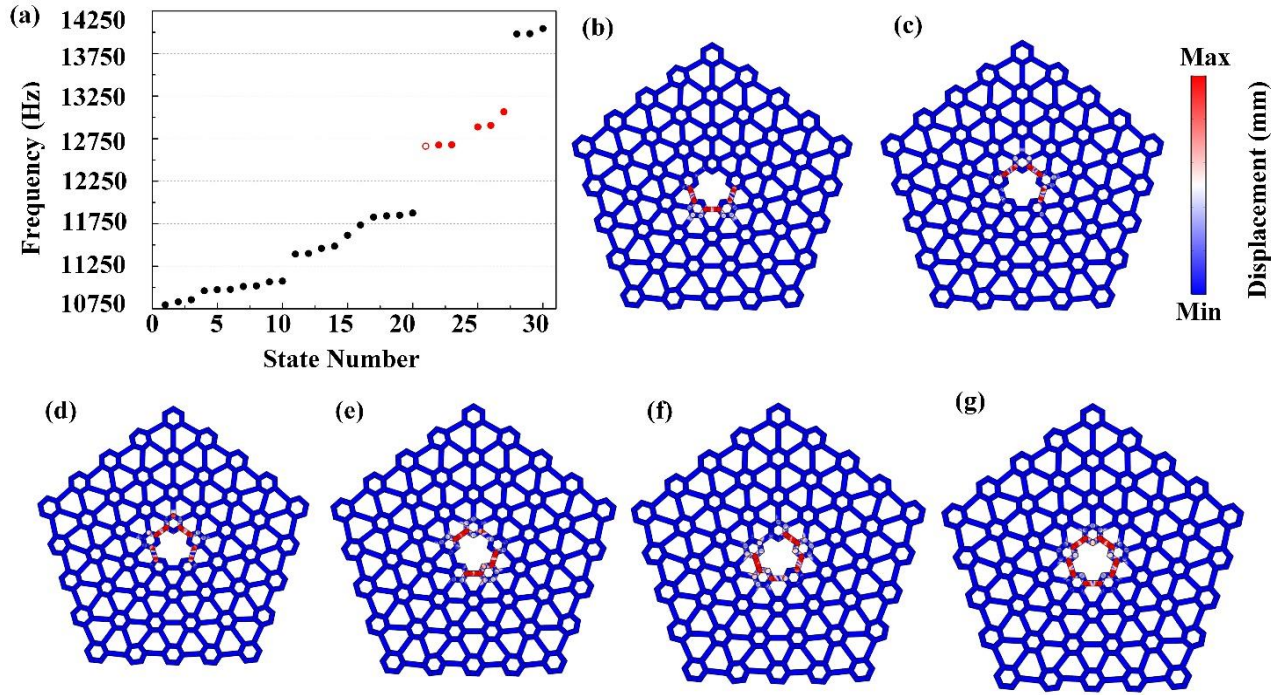

Fig. 14 (a) Eigenfrequency spectrum of the topologically nontrivial elastic phononic plate with the pentagonal core yielded by disclinations. The number of unit cells is 55. The red solid circles are the topological bound states. The red hollow circle is the trivial defect state. The black solid circles are the bulk states. (b) The simulated displacement field profile of the trivial defect state localized around the pentagonal core. (c)-(g) The simulated displacement field profiles of five topological bound states localized around the pentagonal core.

The topological nontrivial elastic phononic plate immunizes against moderate imperfects. By removing several additional masses, the five topological bound states are still observed in the band gap and isolated from the trivial defect states yielded by imperfects (seeing in Fig. 15). For trivial defect states, the elastic wave energy is trapped at the node without additional masses, seeing Fig. 15(c). Furthermore, the trivial bound state, which cannot be observed in the same plate without disturbances (seeing Fig. 12), is generated in this sample. According to eigenfrequency spectra presented in Figs. 12 to 15, we can obtain that the trivial bound state is seriously instable. The finite sizes and the disturbances of plates can make it to be gone and generated. However, the topological bound states are robust and insensitive to the finite sizes and disturbances of plates.

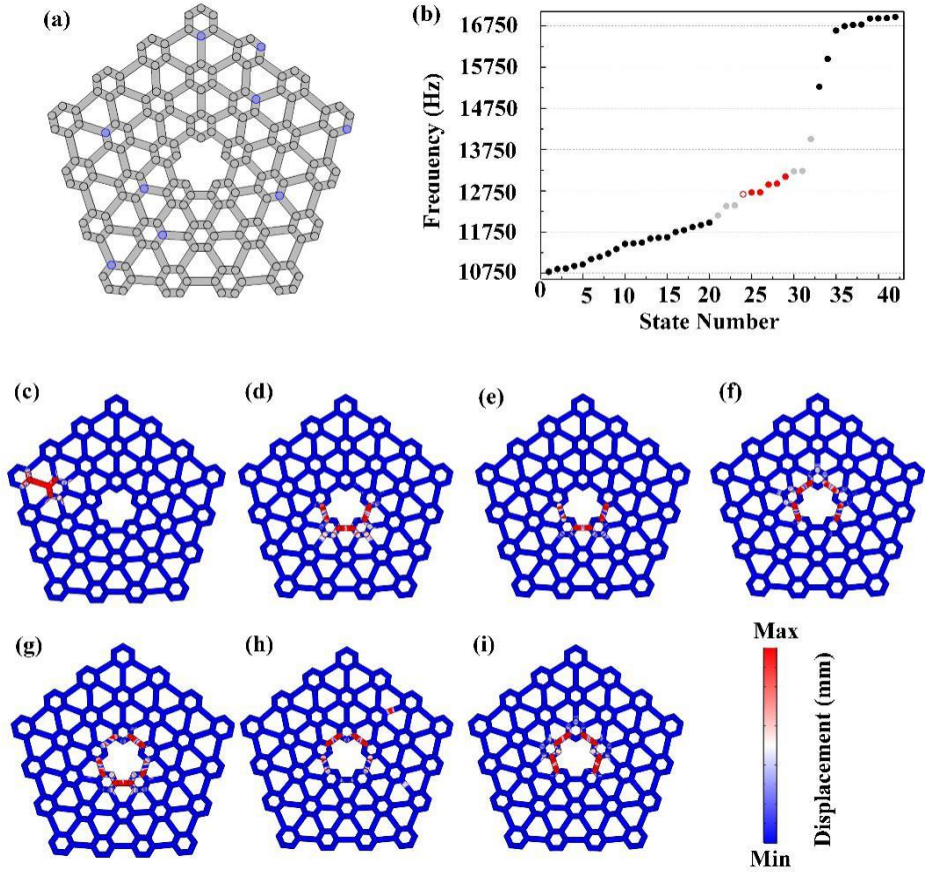

Fig.15 (a) The topologically nontrivial phononic plate with thirty-unit cells. Nodes without additional masses are marked by blue circles. (b) Eigenfrequency spectrum for the topologically nontrivial elastic phononic plate with imperfects. The red solid circles are the topological bound states. The red hollow circle is the trivial defect state. The black solid circles are the bulk states. The grey solid circles are the defect states yielded by removing several additional masses. (c) The simulated displacement field profile of the trivial defect state yielded by imperfect. (d) The simulated displacement field profile of the trivial defect state localized around the pentagonal core. (e)-(i) The simulated displacement field profiles of five topological bound states localized around the pentagonal core.

For the trivial elastic phononic plate, the topological bound states cannot be found and the deformation of the plate will yield lots of trivial bulk states in band gaps. As shown in Fig. 16(b), when the number of unit cells is 15, there is no in-gap bound state and the bulk band gap becomes very narrow due to the deformation of unit cells. As shown in Fig. 16(d), when the number of unit cells is 30, there is still no in-gap bound state and the bulk band gap is almost vanished. Lots of bulk states yielded by the deformation of lattice almost spread out the band gap.

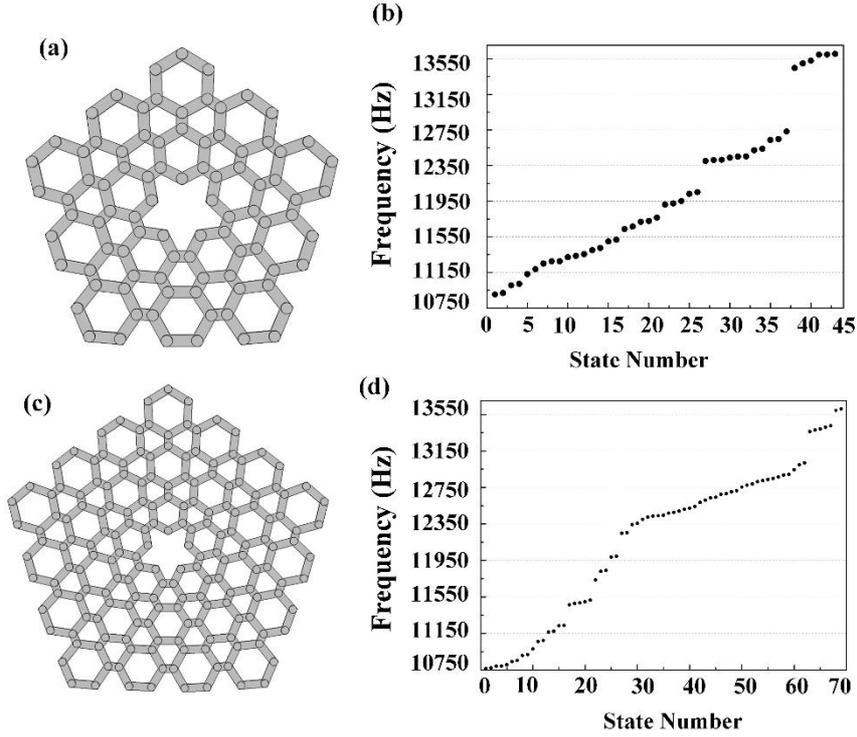

Fig. 16 Trivial elastic phononic plate with the pentagon core. The number of unit cells is 30. (b) Eigenfrequency spectrum for the trivial elastic phononic plate with thirty-unit cells. (c) Trivial elastic phononic plate with the pentagon core. The number of unit cells is 15. (d) Eigenfrequency spectrum for the trivial elastic phononic plate with fifteen-unit cells. The black solid circles are the bulk states.

## *2.3 Bound states of elastic phononic plates without hollow pentagonal cores*

Now, the pentagonal hollow region is filled with a pentagonal unit cell which connects their nearest-neighbor hexagonal unit cells with beams. There are also five topological bound states in the band gap, as shown in Fig. 17(a). The eigenfrequencies of these five topological bound states are almost the same as the former ones without the pentagonal unit cell. Intensity distributions of five topological bound states depicted in Figs. 17(b)-17(f) show that the elastic waves strongly localize around the boundaries of the pentagonal core, without diffusing into the pentagonal unit cell just as it does not exist. It should be noted that the trivial bound mode cannot be observed in this sample as the hollow region is filled by the pentagonal unit cell. For the hexagonal plates without disclinations, there are no bound states in band gaps, both with and without hollow regions (seeing Fig. 18). The reason is that there are six boundaries around the hexagonal core, giving rise to an overall charge of 3.

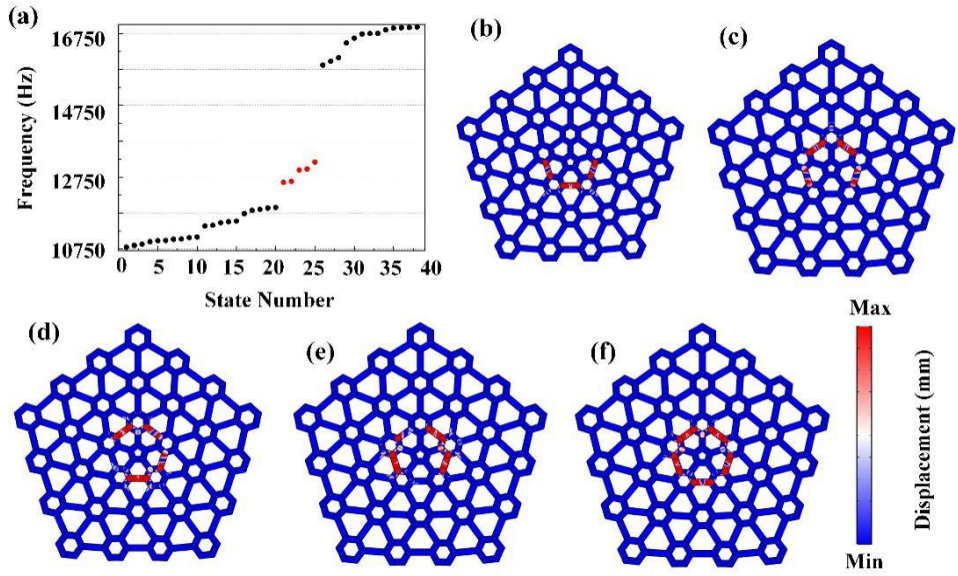

Fig. 17 (a) Eigenfrequency spectrum for the topologically nontrivial elastic phononic plate whose pentagonal core is filled with a pentagonal unit cell. The red solid circles are the topological bound states. The black solid circles are the bulk states. (b)-(f) The simulated displacement field profiles of five topological bound states localized around the pentagonal core.

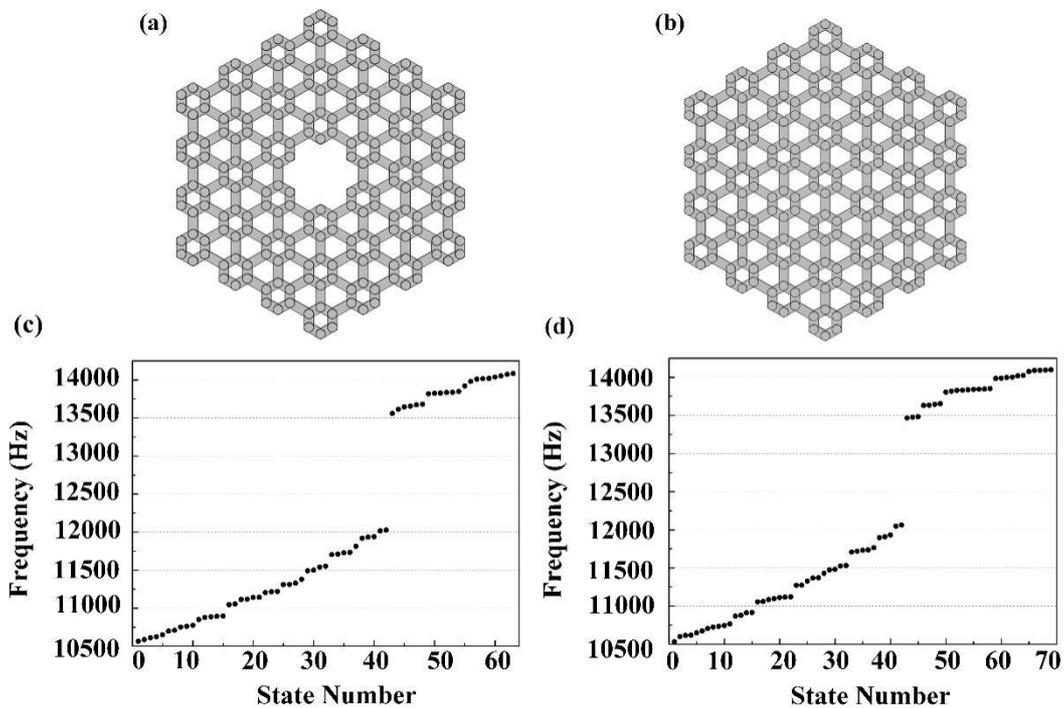

Fig. 18 (a) and (b) Topological nontrivial hexagonal elastic phononic plate with and without the hollow region. (c) and (d) Eigenfrequency spectrum for the topological nontrivial hexagonal elastic phononic plate with and without the hollow region.

# 5 Conclusion

In this paper, we investigated topological defect states in elastic phononic plates. Firstly, we produced the topological bound states induced by disclinations of lattices. The topological bound states immunize against the finite size and the moderate imperfect of the plate. The well localized bound states provide a new platform for the modulation of the elastic wave, and exhibit good application prospects in the high-sensitive sensing and the energy recovery. Then, by inducing disclinations and dislocations into valley-polarized elastic phononic plates, the deformation of lattices yields topologically protected interfaces with greatly robust propagation against cavities and disorders. Essentially differing from general valley-polarized interfaces which are elaborately designed between two distinct lattices with opposite topological charges, our interfaces is yielded in a same lattice whose topological phase transition across interfaces is yielded by the deformation of lattices. From both cases, the deformation of lattices induced by topological defects provides a new degree of freedom to follow the desired waveguide routes and bound states. Though they are implemented in elastic wave systems, these topological defect states can be developed to other wave systems, such as acoustic, optical, electromagnetic and electric systems, for localized polarization and wave guiding.

**Acknowledgements**

This work is supported by the National Natural Science Foundation of China (Grant No. 12072108, 51621004).